\documentclass[%
 reprint,
 jmp,
superscriptaddress,
 amsmath,amssymb,
 aps,
]{revtex4-2}

\usepackage{graphicx}
\usepackage{dcolumn}
\usepackage{bm}
\usepackage{mathtools}

\usepackage{braket}
\usepackage{xcolor}
\usepackage{bbold}
\begin{document}

\preprint{APS/123-QED}
\title{Quantum interference with time-frequency modes and multiple-photons generated by a silicon nitride microresonator} 

\author{Massimo Borghi}
\email{corresponding author: massimo.borghi@unipv.it} 
\author{Emanuele Brusaschi}
\author{Marco Liscidini}
\author{Matteo Galli}
\affiliation{Dipartimento di Fisica, Università di Pavia, Via Bassi 6, 27100 Pavia, Italy.}
\author{Daniele Bajoni}
\affiliation{Dipartimento di Ingegneria Industriale e dell'Informazione, Universit\`a di Pavia, Via Adolfo Ferrata 5, 27100 Pavia, Italy.}


\date{\today}

\begin{abstract}
We demonstrate bipartite gaussian boson sampling with squeezed light in $6$ mixed time-frequency modes. Non-degenerate two-mode squeezing is generated in two time-bins from a silicon nitride microresonator with simultaneous high spectral purity ($>0.86(3)$) and indistinguishability ($0.985(2)$). An unbalanced interferometer embedding electro-optic modulators, which is stabilized by exploiting the continuous energy-time entanglement of the generated photon pairs, controls time and frequency-bin modes. 
We measure $144$ collision-free events with $4$ photons at the output, achieving a fidelity $>0.98$ with the theoretical probability distribution. We use this result to identify the similarity between families of isomorphic graphs with $6$ vertices, and present an approach for the realization of universal operations on time-frequency modes.\\

\end{abstract}

\maketitle


\section{\label{sec:intro} Introduction}
Most of the protocols for quantum computation and communication rely on the interference of multiple and identical photons. For example, the generation of multi-partite entangled states for measurement based quantum computation, such as three-photon Greenberger-Horne-Zeilinger states,  can be achieved by feeding multiple photons to a network of beasmplitters and phase shifters, and by heralding on specific photodetection patterns at the output  \cite{gimeno2015three,krenn2017quantum,paesani2021scheme,cao2024photonic,bao2023very}.
Quantum teleportation and entanglement swapping are based on a Bell state measurement, which is ultimately linked to the Hong Ou Mandel (HOM) interference between identical photons \cite{azuma2023quantum}. At a more fundamental level, generalized HOM tests with multiple-photons allows one to infer collective properties of a quantum state, such as the $n$-photon indistinguishability \cite{pont2022quantifying}, or to discern genuine multi-photon interference from alternative models of reduced computational complexity \cite{tillmann2015generalized,giordani2018experimental}.\\
Specialized instances of multi-photon interference are boson-sampling problems\cite{kruse2019detailed}.
The complexity of sampling from the probability distribution of multiple photons at the output of a linear multiport interferometer is inherent to the bosonic nature of photons. Indeed, the exchange symmetry of identical bosons leads to a probability distributions that are functions of the properties of the scattering matrix describing the interferometer, such as the permanent or the Hafnian, which are hard to compute with the best-known classical algorithms \cite{kruse2019detailed}. 
A variant of Aaronson and Arkhipov's original protocol of boson sampling \cite{aaronson2011computational}, which uses gaussian states at the input instead of single photons, has gained increasing attention because it has eliminated the necessity of generating $n$ indistinguishable photons on-demand, greatly improving the scale of experiments and detection rates \cite{GBS_invenzione,kruse2019detailed}.  For these reasons, gaussian boson sampling (GBS) has already demonstrated quantum computational advantage for specific tasks \cite{Quant_adv_xanadu,Quantum_adv_Pan,Quantum_adv_Pan_2}. The interest in GBS is also motivated by the numerous applications where it might provide a speed-up over classical machines. In particular, when squeezed light is fed at the input of an interferometer, one can relate the output distribution to the number of perfect matching on a graph \cite{GBS_application}.  
This connection has inspired several theoretical proposals for graph optimization \cite{Graph_opt_1, Graph_opt_2, Graph_opt_3}, simulation of molecular docking \cite{Molecular_docking}, evaluation of graph similarity and description of point processes \cite{Point_processes}. The propagation of squeezed light in an interferometer is described by a sequence of gaussian operations that can also be exploited to perform several different tasks including the calculation of the vibronic spectra of molecules \cite{Vibronic_spectra,Sampling_Paesani,zhu2024large}.\\
Most of the GBS experiments reported to date are based on path or time-bin encoding of optical modes, which are manipulated through bulk \cite{Quant_adv_xanadu,Quantum_adv_Pan,Quantum_adv_Pan_2,yu2023universal} or integrated photonic setups \cite{Sampling_Paesani,arrazola2021quantum,zhu2024large}. Additionally,  the combination of two different degrees of freedom (e.g. path and polarization \cite{Quantum_adv_Pan}), can be exploited to scale the size of the experiment.\\   
One degree of freedom  which has so far remained  relatively unexplored in sampling experiments is  frequency-bin. The interference of two-photons in two discrete frequency-bins has been experimentally demonstrated by many groups using electro-optic modulators \cite{imany2020probing,imany2019high,reimer2016generation,khodadad2021spectral,imany2018frequency}, Bragg scattering four-wave mixing  (BSFWM) \cite{joshi2020frequency,lee2024noon,li2019tunable} or difference frequency generation \cite{kobayashi2016frequency}. Recently,  BSFWM has been used to interfere two photons into three frequency modes by using multiple pumps and a careful dispersion engineering of the nonlinear medium where they interact \cite{oliver2024n}. Few experiments involved more than two photons in the frequency-bin domain. For example, in \cite{reimer2016generation} frequency was used as an auxiliary mode to generate four-photon entangled states in the time-bin domain, while the frequency-resolved four-photon emission from a photon pair source was used in \cite{bell2020diagnosing} to diagnose the phase correlations in the joint spectral amplitude, or in ref.\cite{bell2018multiphoton} to measure the interference between concurring nonlinear optical processes. However, there are several aspects which make frequency-bin encoding attractive for sampling experiments. For example, a very large number of modes can be defined in a dense wavelength grid and processed in parallel in a single spatial mode and with standard fiber-optic components \cite{lu2023frequency,henry2023parallelizable,henry2024parallelization}. Moreover, the inherent high-dimensionality of frequency-bin encoding can be further increased by combining it with time \cite{imany2019high,reimer2019high}, polarization \cite{wen2023polarization,miloshevsky2024cmos}, and path encoding \cite{zhang2023chip}. \\
Following these ideas, we propose and demonstrate
a bipartite GBS experiment in which photons are sampled in six time-frequency-bin modes \cite{grier2022complexity,arrazola2021quantum}.  Two-mode squeezed (TMS) light is generated by an integrated silicon nitride microresonator by spontaneous four-wave mixing at not degenerate wavelengths and in two sequential time-bins. The high spectral purity of the photons within each TMS and the almost perfect indistinguishability between the two squeezers is assessed through their joint temporal intensity (JTI) \cite{borghi2024uncorrelated} and the visibility of the HOM interference between photons heralded from different temporal modes \cite{faruque2018chip}. Squeezed light is sent into an unbalanced interferometer embedding electro-optic phase modulators that scramble the time and frequency-bin modes.
We focus on the measurement of $144$ four-photon detection patterns at the output, and reconstructed their probabilities, which are in close agreement with those predicted by simulations. We provide evidence that samples arise from squeezed light fed at the input of the interferometer by adopting a Bayesan test that rules out alternative models based on classical light \cite{Sampling_Paesani}. 
The output samples are used to construct feature vectors that are exploited to cluster bipartite-graphs based on their similarity \cite{BGS_orbite,GBS_application,arrazola2021quantum}. Finally, we propose a scheme to perform arbitrary operations on time-frequency modes, a necessary but still missing ingredient that would promote time-frequency modes as a competitive alternative to the current schemes of GBS based on path and time-bin encoding. 

\section{\label{sec: Setup e photon characterization} Experimental setup and characterization of the microresonator source}
\begin{figure*}
    \includegraphics[width = 1 \textwidth]{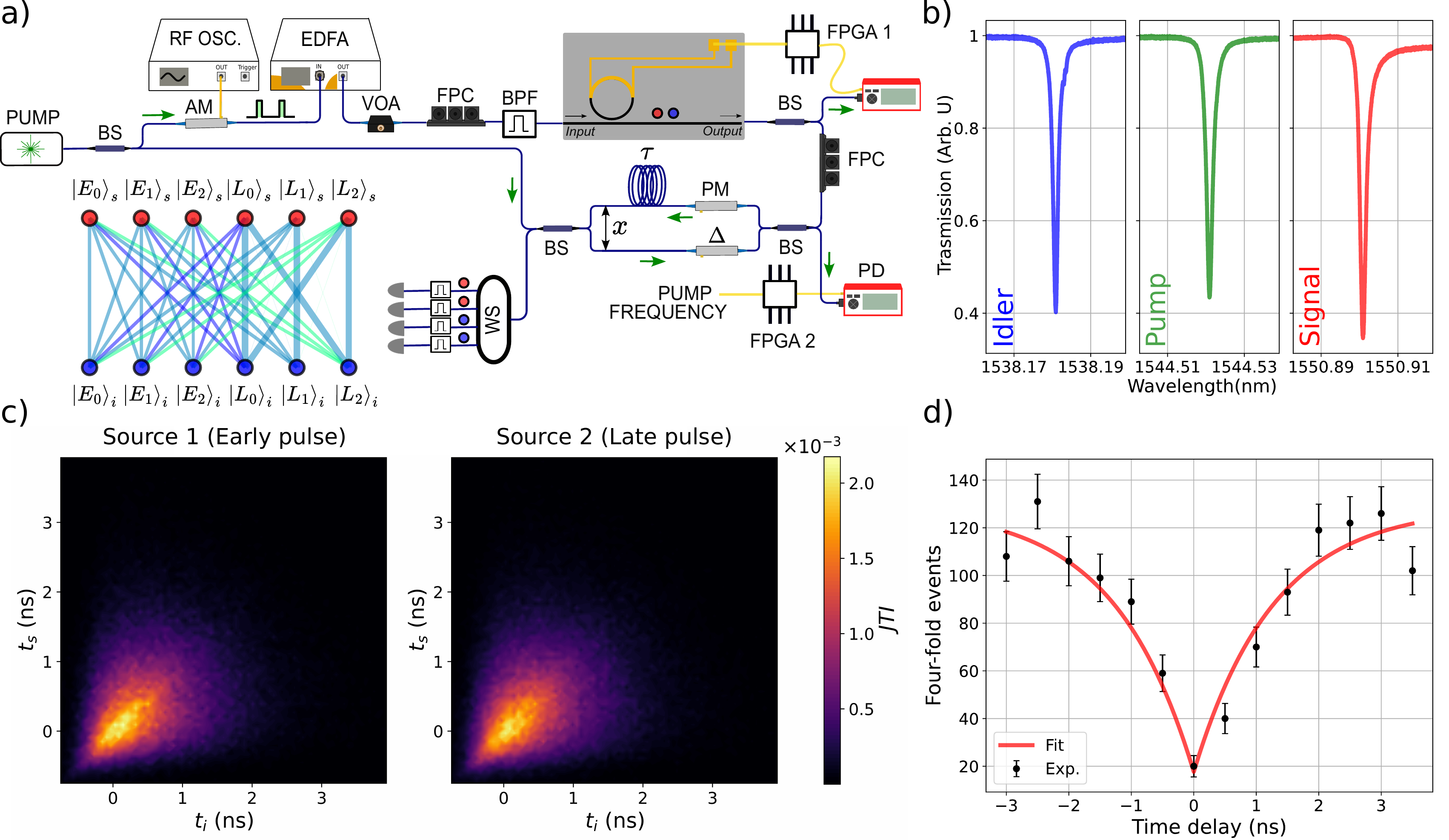}
    \caption{(a) Sketch of the experimental setup. The pump laser is shown in green, while signal and idler modes are represented with a red and blue circle respectively. The inset is an illustrative example of a graph that can be encoded in the GBS experiment. The width of the edges represents the module of the complex weight, while the color encodes its phase. BS: beamsplitter, VOA: variable optical attenuator, BPF: bandpass filter, FPC: fiber polarization controller, EDFA: Erbium doped fiber amplifier, FPGA: Field Programmable Gate Array, PM: phase-modulator, WS: waveshaper, PD: photodiode, AM: amplitude modulator. (b) Normalized transmission spectra of the pump, signal and idler resonances. (c) Joint temporal intensity of photon pairs generated in the early and late pump pulses. (b) Four-fold coincidences (black) in the heralded HOM experiment as a function of the time-delay between the two heralded photons. The solid red line is a fit of the data which uses Eq.(\ref{fit_HHOM}).
    }
    \label{Fig_0}
\end{figure*}
\subsection{Experimental setup \label{subsec:experimental_setup}}
The experimental setup is shown in Fig.\ref{Fig_0}(a). A butterfly laser diode emitting at a wavelength of $1544.53$ nm is sent to an amplitude electro-optic modulator (EOM) to carve two sequential rectangular pulses with a time duration of $800$ ps, a delay $\tau=20$ ns, and a repetition rate of $20$ MHz. Besed on their time of emission with respect to an electronic trigger carrying the $20$ MHz clock, the two pulses define the early (E) and the late (L) temporal modes. 
The pulses are amplified by an Erbium-doped fiber amplifier (EDFA) and the background noise is filtered by a passband filter. A fiber polarization controller is used to set the polarization to TE before coupling light to the chip by using an Ultra-High Numerical Aperture (UNHA7) fiber. The optimized coupler and the use of index-matching gel reduce the coupling loss to $\sim 1$ dB/facet and ensure mechanical stability ($<0.2$ dB of drift during $24$ hours of continuous measurement). The pump light is sent to a silicon nitride microresonator, generating bichromatic Two-mode squeezed states(TMS) by Spontaneous Four-Wave Mixing(SFWM) over multiple pairs of resonances symmetrically located in frequency with respect to the pump. The resonator has a waveguide cross-section of $1.5\times 0.8\,\mu$m and a perimeter of $835\,\mu$m, corresponding to a free spectral range of $\sim200$ GHz. The resonator is over-coupled to the bus-waveguide, with a loaded quality factor of $Q\sim 8\times10^5$ and a measured escape efficiency of $\sim 0.75$. The pump resonance is tuned to the wavelength of the input laser using a thermal phase shifter placed on the top of the resonator. The signal and idler modes used in the experiment are at a wavelength of $\lambda_{s,1} = 1550.9$ nm ($\nu_{s,1}=193.3$ THz) and $\lambda_{i,1} = 1538.18$ nm ($\nu_{i,1}=194.9$ THz). The spectra of these resonances is shown in Fig.\ref{Fig_0}(b). By assuming nearly single-mode spectral emission (the assumption is justified in Section \ref{sub:characterization_source}), the state $\ket{\Psi}$ at the output of the chip can be written as
\begin{equation}
\ket{\Psi} = \left (e^{\xi_{E_1}a^{\dagger(s)}_{E_1}a^{\dagger(i)}_{E_1}+\xi_{L_1}a^{\dagger(s)}_{L_1}a^{\dagger(i)}_{L_1}-\textrm{h.c.}}\right )\ket{\textrm{vac}},
\label{eq:state_input}
\end{equation}
where $a^{\dagger(s/i)}_{E/L,1}$ is the creation operator of the signal/idler photon in the early/late temporal mode at 
the resonance frequency $\nu_{s/i,1}$. $\xi_{E_1}$ and $\xi_{L_1}$ are the corresponding squeezing parameters. In our case the two pump pulses have equal intensity, and their time separation is sufficiently long to reset the state of the resonator between the pulses. Therefore, one can set $|\xi_{E_1}|=|\xi_{L_1}|=|\xi|$. However, the relative phase between the two terms in Eq.(\ref{eq:state_input}) can be different.\\
Light at the output of the chip is sent to a $99/1$ beamsplitter, and we used the $1\%$ tap to actively lock the resonance wavelength to the pump laser using an Field Programmable Gate Array (FPGA) that controls the heater on the resonator. The transmitted light is coupled to an interferometer that allows one to manipulate the state. Its architecture is a fiber-based Mach-Zender interferometer \cite{Franson} in which an electro-optic phase modulator (PM)  is inserted in both arms. A $\sim 4$m-long fiber introduces a time delay of $20$ ns between the two paths, matching the time separation between the pump pulses. The unbalanced interferometer is used to mix the temporal modes, while the PMs, driven by a $18.2$ GHz tone and with a modulation index of $\delta\sim1.4$, scramble the frequency modes. The relative phase $x$ between the long and short arm of the interferometer can be controlled, and it is stabilized using the pump laser as reference in an active feedback loop implemented via a second FPGA. Due to the time-energy entanglement of the photon pairs generated in the SFWM process, the locking of the pump phase stabilizes the joint relative phase of the signal and the idler photon between the long and the short path of the interferometer . \\
At the output of the interferometer, the pump is suppressed by a band-pass filter of high-rejection, and a programmable waveshaper is used to select the frequencies of the signal and idler photons that are detected.
The detection is performed using four superconducting nanowire single-photon detectors (SNSPD) with a detection efficiency of $85\%$. 

\subsection{Characterization of the photon source \label{sub:characterization_source}}
High-visibility multi-photon interference requires indistinguishable and pure photons \cite{Drago_HOM_24,faruque2018chip}. When photon pairs are generated by SFWM in a microresonator, high spectral purity is obtained when the pump pulse has a temporal duration larger than the dwelling time of light in the cavity. From the loaded quality factor $Q$ and the pump frequency $\nu_p$, the estimated dwelling time is $Q/\nu_p\sim660$ ps. We work with a pulse duration of $800$ ps, which optimizes the pulse shape  carved by the amplitude EOM. The characterization of the spectral purity of the photons within each TMS is done by measuring the joint temporal intensity (JTI) \cite{borghi2024uncorrelated}. 
The JTI is reconstructed by measuring the arrival times of the signal and idler photons with respect to the $20$ MHz electronic trigger and by binning the timestamps into a two-dimensional grid with a resolution of $70\,\textrm{ps}\times70\,\textrm{ps}$, which is limited by the jitter of our detectors. The JTIs of the early and late temporal modes are measured simultaneously to discern any distinguishability between photons generated in the first or in the second pulse.
The two JTIs are reported in Fig.\ref{Fig_0}(c). Their shape has been already discussed in \cite{borghi2024uncorrelated}. The overlap between the two distributions is an upper bound to the indistinguishability, and is $0.985(2)$. From the JTI it is also possible to determine an upper bound to the purity $\mathcal{P}$ of the generated photons, which is $0.928(5)$ for those of the early pulse and $0.93(1)$ for those of the late pulse \cite{borghi2024uncorrelated}. This value is very close to the theoretical limit achievable with a resonator source having signal, pump and idler resonances with the same quality factor, and without pump engineering \cite{vernon2017truly}. \\
The purity is also estimated by performing HOM interference between photons heralded in different temporal modes.
In this experiment, the idler photons at the output of the chip are sent to a $50/50$ beasmplitter, while the signals are sent to the interferometer, and detected at the two output ports. 
The idlers herald two signal photons at the input of the interferometer with a relative delay $\tau$. By post-selecting the events in which two signal photons exit the two output ports of the interferometer in the late time-bin, we select the cases in which they arrived at the output beamsplitter from different ports and at the same time, a scheme of the experiment is report in Methods Fig. \ref{HHOM_scheme}(a). 
To assess the visibility of the HOM dip, we varied the delay $\tau'$ between the first and the second pulse without modifying the time delay of the interferometer. This introduces a time difference $\Delta\tau= \tau'-\tau$ in the arrival of the two photons at the output beamsplitter. 
In order to mitigate the emission of multiple pairs from each source, the pump power is regulated to achieve a pair generation probability per pulse $\sim 0.03$.
The number of four-fold coincidences $C$ as a function of $\Delta \tau$ is shown in Fig.\ref{Fig_0}(d). We fit the data using the model equation
\begin{equation}\label{fit_HHOM}
C = B (1-V e^{-|\Delta \tau|/\tau_c}),
\end{equation}
where $V$ is the visibility of the HOM dip, $\tau_c$ is the resonator dwelling time and $B$ is the maximum rate of four-fold detections. The choice of Eq.(\ref{fit_HHOM}) is motivated by the lorentzian lineshape of the signal/idler resonances (which is a decaying-exponential in the time-domain) and by the use of a rectangular pump pulse to trigger the SFWM process. The raw visibility extracted from the fit is $0.86(3)$. 
The visibility of the HOM dip is mainly limited by thermal-noise and by the spectral purity of the heralded photons (see Methods section). 
We also performed the HOM interference between photons heralded from different pulses when their frequency is changed by the PMs before interfering at the output beasmplitter. The frequency of the photon in the early(late) pulse is changed by the PM placed in the long(short) of the interferometer. The two frequencies are $\nu_{s/i,0} = \nu_{s/i,1}-\Delta\nu$ and $\nu_{s/i,2} = \nu_{s/i,1}+\Delta\nu$, with $\Delta\nu = 18.2$ GHz. We measured a HOM visibility of $0.8(1)$ for $\nu_{s/i,0}$ and $0.91(5)$ for $\nu_{s/i,2}$, which confirms that the two PMs do not change the Lorentzian spectra of the heralded photons when they are scattered into the same frequency-bin.

\section{Bipartite GBS experiment}
\label{sec:multi-photon_interference}
The state in Eq.(\ref{eq:state_input}) is sent at the input of the interferometer shown in Fig.\ref{Fig_0}(a), which allows us to couple both time and frequency modes. 
The PMs inserted in the arms of the interferometer can scatter photons to different frequencies $\nu_{s/i,j} = \nu_{s/i,1}+(j-1)\Delta\nu$ with an efficiency $|J_{j-1}(\delta)|^2$, where $J_j(\delta)$ is the Bessel function of the first kind of order $j$ evaluated at the modulation index $\delta$.  For the moderately low modulation index used in the experiment ($\delta\sim 1.4$), most of the energy is scattered into three frequency modes, which are $j={0,1,2}$. \\Events for which photons are scattered out of these frequencies are discarded. The six modes that are coupled by the interferometer can be labeled as $\ket{E(L)_j}_{s(i)}$, where $E(L)$ labels the temporal bin, Early(E) and Late(L), and $j=(0,1,2)$ the frequency bin, where we order the basis as $\{ \ket{E_0},\ket{E_1},\ket{E_2},\ket{L_0},\ket{L_1},\ket{L_2}\}$. Four-photon events consisting of two signal and two idler photons are post-selected at the output of the interferometer (see Fig.\ref{Fig_0}(a)).   
In general, there are $\binom{N+k-1}{k}$ distinct ways in which $k$ identical photons can be arranged in $N$ modes. However in our work, since signal and idler photons are distinguishable, we have for two indistinguishable signals $k=2$ and $N=6$, and the same for the idler, which yields $\binom{N+k-1}{k}^2=441$ possible output patterns. However, not all of them can be observed in the experiment. The use of threshold detectors implies that only collision-free events, with no more than one photon in each mode, can be distinguished. Moreover, given that the time-bin separation is smaller than the dead-time of the detectors ($\sim 80$ ns), only one temporal mode can be detected for each frequency. The observable combinations have the form $\ket{X_mY_n}_s\ket{X'_pY'_q}_i$, with $X,X'$ and $Y,Y'$ $\in \{E,L\}$ and $m,n,p,q$ $\in \{0,1,2\}$ with $m\neq n$ and $p\neq q$.
Out of the $441$ combinations, those which satisfy these requirements are $144$.\\
To calculate the probability of observing a particular output pattern $S=(X_m,Y_n,X_p,Y_q)$, we map the experiment into an instance of bipartite gaussian boson sampling \cite{grier2022complexity}, in which two-mode squeezers over $N$ modes are sent at the input of an interferometer described by a $2N\times2N$ complex matrix $T_{si}=T_s\bigoplus T_i$, where $T_{s(i)}$ is a $N\times N$ matrix describing the transformation of the interferometer on the signal(idler) photon. The probability of observing $S$ is given by \cite{arrazola2021quantum}
\begin{equation}
p(S) = \frac{\textrm{Perm}(C_s)|^2}{\sqrt{\textrm{det}(\sigma_Q)}}\label{Haf_to_perm_form},
\end{equation}
where $\textrm{det}(\sigma_Q)=\cosh^4{\xi}$ \cite{kruse2019detailed} and $C_s$ can be formed from the matrix $C$
\begin{equation}
    C = T_i \Bigl( \bigoplus_{i = 1}^m \textrm{tanh}(\xi_i)  \Bigl) T_s^T. \label{eq:Cmatrix}
\end{equation}
from the intersections between the rows and columns of $C$ corresponding respectively to the signal and idler modes which are detected. For example, if $S=(E_0,L_1,E_2,L_2)$, the ordering of the modes introduced before implies that rows $(1,4)$ and columns $(3,6)$ are selected from $C$ to form $C_s$. In writing Eq.(\ref{Haf_to_perm_form}) we implicitly assumed that no more than one photon is detected in each mode. It is worth to stress that Eq.(\ref{Haf_to_perm_form}) 
holds only when the modes $S'\neq S$ are post-selected to be all in vacuum state and lossless photon number resolving detection is performed over the $S$ modes \cite{kruse2019detailed}. However, as shown in the Methods section, for the moderate squeezing strength of the experiment, Eq.(\ref{Haf_to_perm_form}) provides an excellent approximation to the exact output probability.  Equation \ref{Haf_to_perm_form} can be rewritten as 
\begin{equation}
    p(S) = \frac{|\textrm{Haf}(\mathcal{A}_s)|^2}{\sqrt{\textrm{det}(\sigma_Q)}}, \label{eq:hafnian}
\end{equation}
where $\textrm{Haf}(\mathcal{A})$ denotes the Hafnian of the matrix 
\begin{equation}
    \mathcal{A}_s = 
    \begin{bmatrix}
        0 & C_s \\
        C_s^T & 0
    \end{bmatrix} ,\label{eq:adj_matrix}
\end{equation}
By using the well known relation between the calculation of the Hafnian and perfect matchings on graphs \cite{kruse2019detailed}, the matrix $\mathcal{A}_s$ can be interpreted as the complex adjacency matrix of a bipartite graph, where the two vertex sets are formed by respectively the signal and the idler modes. As it will be shown later, the graph associated to the experiment is a complete bipartite graph. An example is shown in Fig.\ref{Fig_0}(a) for a particular setting of the PMs and interferometer phases. \\
The matrices $T_s$ and $T_i$ describing the action of the interferometer 
on the time-frequency modes have the general form
\begin{equation}\label{Matrice_idler}
    T_{s(i)} = \frac{1}{2}
    \begin{bmatrix}
        \cdot & J^{E}_{-1} & \cdot & \cdot & \cdot & \cdot\\
        \cdot & J^{E}_{0} & \cdot & \cdot & \cdot & \cdot\\
        \cdot & J^{E}_{1} & \cdot & \cdot & \cdot & \cdot\\
        \cdot & J^{L}_{-1} & \cdot & \cdot & J^{E}_{-1}e^{i \theta^{s(i)}_{1}} & \cdot\\
        \cdot & J^{L}_{0} & \cdot & \cdot & J^{E}_{0}e^{i \theta^{s(i)}_{2}} & \cdot\\
        \cdot & J^{L}_{1}& \cdot & \cdot & J^{E}_{1}e^{i \theta^{s(i)}_{3}} & \cdot
    \end{bmatrix},
\end{equation}
where $J^{E/L)}_i$ is the Bessel function of the first kind of order $i$ describing the strength of the electro-optic modulation of the PM inserted in the short (E) and long (L) arm of the interferometer. The matrix entries denoted with dots are not relevant for calculating the outcome probabilities because photons enter into the interferometer only at frequencies  $\nu_{s/i,1}$. The factor $1/2$ in Eq.(\ref{Matrice_idler}) accounts for the two $50/50$ beamplitters forming the interferometer. According to \cite{rahimi2013direct}, two matrices $M$ and $M'$ describe the same physical  interferometer if they are related by $M = D_1M'D_2$, where $D_1$ and $D_2$ are diagonal matrices describing pure phase shifts on the modes. This feature allows one to set the entries in a particular row and column, corresponding to a reference mode, to real numbers. We chose the second mode as reference, which allows us to write all the entries of the second column as real numbers. 
The phases $\theta^{s(i)}_j$ physically correspond to the phase difference between the short and long path of the interferometer for the signal(idler) photon whose frequency has been changed by the PM from $\nu_{s/i,1}$ (before entering the interferometer) to $\nu_{s/i,j}$ (after leaving the interferometer). These phases also embed the relative phase between the two TMS. \\
While the determination of $J_i$ can be done straightforwardly by injecting a laser at the input of the two PM and by measuring the intensity of the $i^{th}$ sideband at the output, the determination of the phases $\theta^{s(i)}_j$ require further considerations. In particular, since the interferometer is locked to a fixed value of the phase of the pump $\theta_p$ and not to an absolute reference, $\theta^{s(i)}_1$ fluctuates over time but $\theta^s_1+\theta^i_1=2\theta_p$ is fixed by the time-energy entanglement of the photon-pair \cite{Franson}.  Similarly, signal and idler photons that have changed their input frequencies $\nu_{s(i),1}$ to $\nu_{s,j} = \nu_{s,1}+(j-1)\Delta\nu$  and $\nu_{i,k} = \nu_{i,1}+(k-1)\Delta\nu$ after the PMs will jointly behave as pump photons with frequency $\nu_{p,1}+\Delta\nu \left ( \frac{j+k}{2}-1\right )$. This implies that $\theta^s_j+\theta^i_k=2\theta_p+2\left (\frac{j+k}{2}-1 \right )\Delta$, where $\Delta$ is a phase which depends on $\Delta\nu$ and on the relative phase between the two RF signals driving the two PMs.   
Following these arguments, we can insert Eq.(\ref{Matrice_idler}) into Eq.(\ref{eq:Cmatrix}) and write the matrix $C$ as

\begin{equation}
    C = \frac{\bar{J}^2}{4}
    \begin{bmatrix}
        \mathbb{1}_3 & \mathbb{1}_3 \\
        \mathbb{1}_3 & C_{LL} 
    \end{bmatrix}, \label{eq:Cmatrix_block}
\end{equation}
where $\mathbb{1}_3$ is a $3\times 3$ matrix filled by ones and the block $C_{LL}$ is given by
\begin{equation}
C_{LL} = 
    \begin{bmatrix}
    1 + e^{i(x-2\Delta)}& 1 + e^{i(x-\Delta)} & 1 + e^{ix} \\
     1 + e^{i(x-\Delta)}& 1 + e^{ix} & 1 + e^{i(x+\Delta)} \\
    1 + e^{ix}& 1 + e^{i(x+\Delta)} & 1 + e^{i(x+2\Delta)}
    \end{bmatrix}, \label{eq:block_CLL}
\end{equation}
where $x=2\theta_p+\textrm{arg}(\xi_{L_1}/\xi_{E_1})$ and we have set $J_i^{E/S} = \bar{J}$ for $i=(0,1,2)$. This condition is achieved by the particular choice of the modulation index $\delta=1.4$ in the experiment. 
Hence, from Eq.(\ref{Haf_to_perm_form}) and Eqs.(\ref{eq:Cmatrix_block},\ref{eq:block_CLL}) one has that the probability of detecting a signal-idler pair at modes $q$ and $p$ is given by $|C_{qp}|^2$. 
From Eqs.(\ref{eq:Cmatrix_block},\ref{eq:block_CLL}) one can see that the output of the interferometer only depends on two parameters, which are $x$ and $\Delta$.
The parameter $x$ can be tuned by changing the pump phase at which the interferometer is locked. The parameter $\Delta$ can be modified by varying the phase offset between the RF signals driving the PMs.
\begin{figure}
    \includegraphics[width = 0.5 \textwidth]{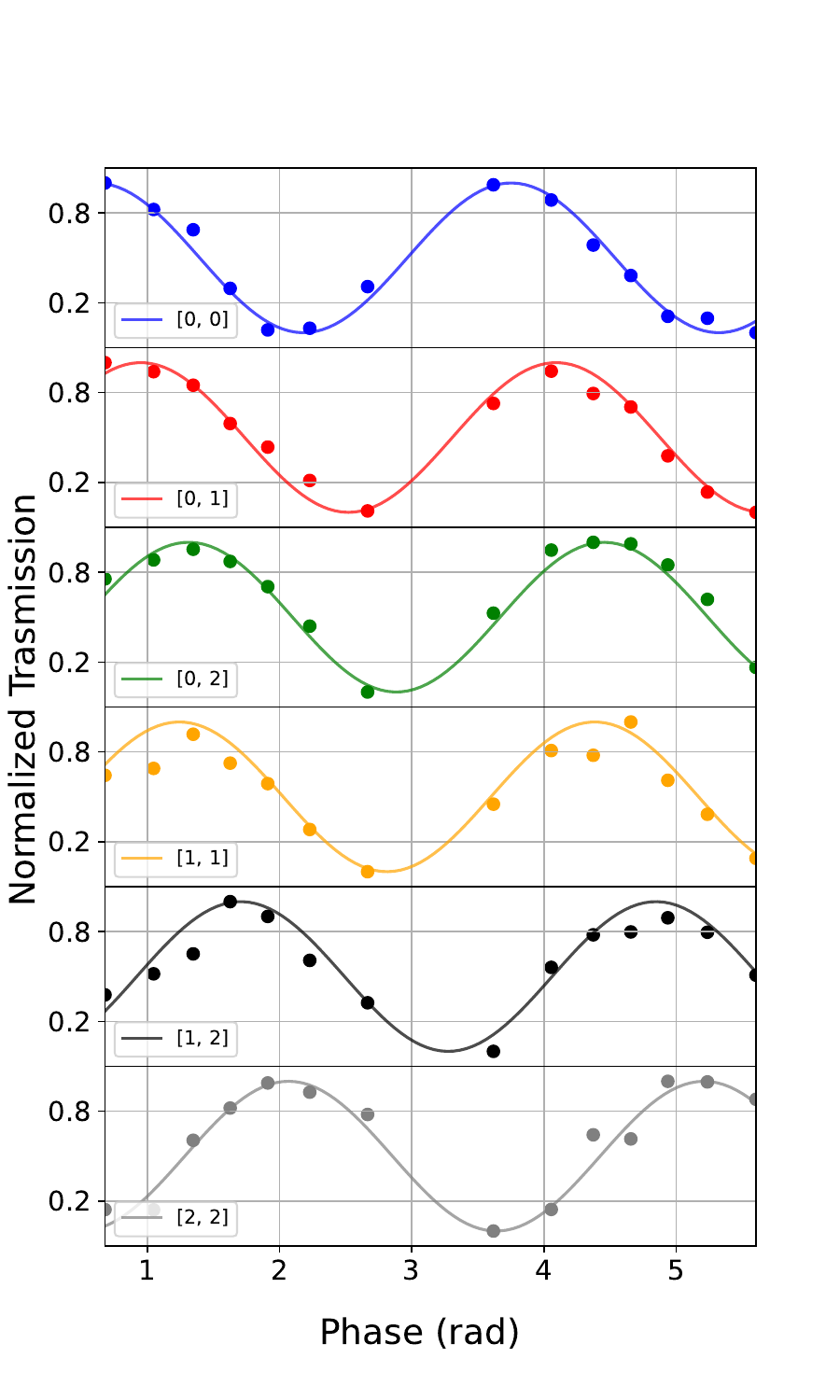}
    \caption{(a) Coincidence probabilities $p_{jk}$ between the signal and the idler photons at frequencies \mbox{$[j,k]=[\nu_{s,1}+(j-1)\Delta\nu,\nu_{i,1}+(k-1)\Delta\nu]$} detected in the late time-bin as a function of the interferometer phase. The datasets have been normalized for clarity. Solid lines are fit of the data which use Eq.(\ref{eq:cos_fit})}
    \label{Fig_Franson}
\end{figure}
\begin{figure*}
    \includegraphics[width = 0.9 \textwidth]{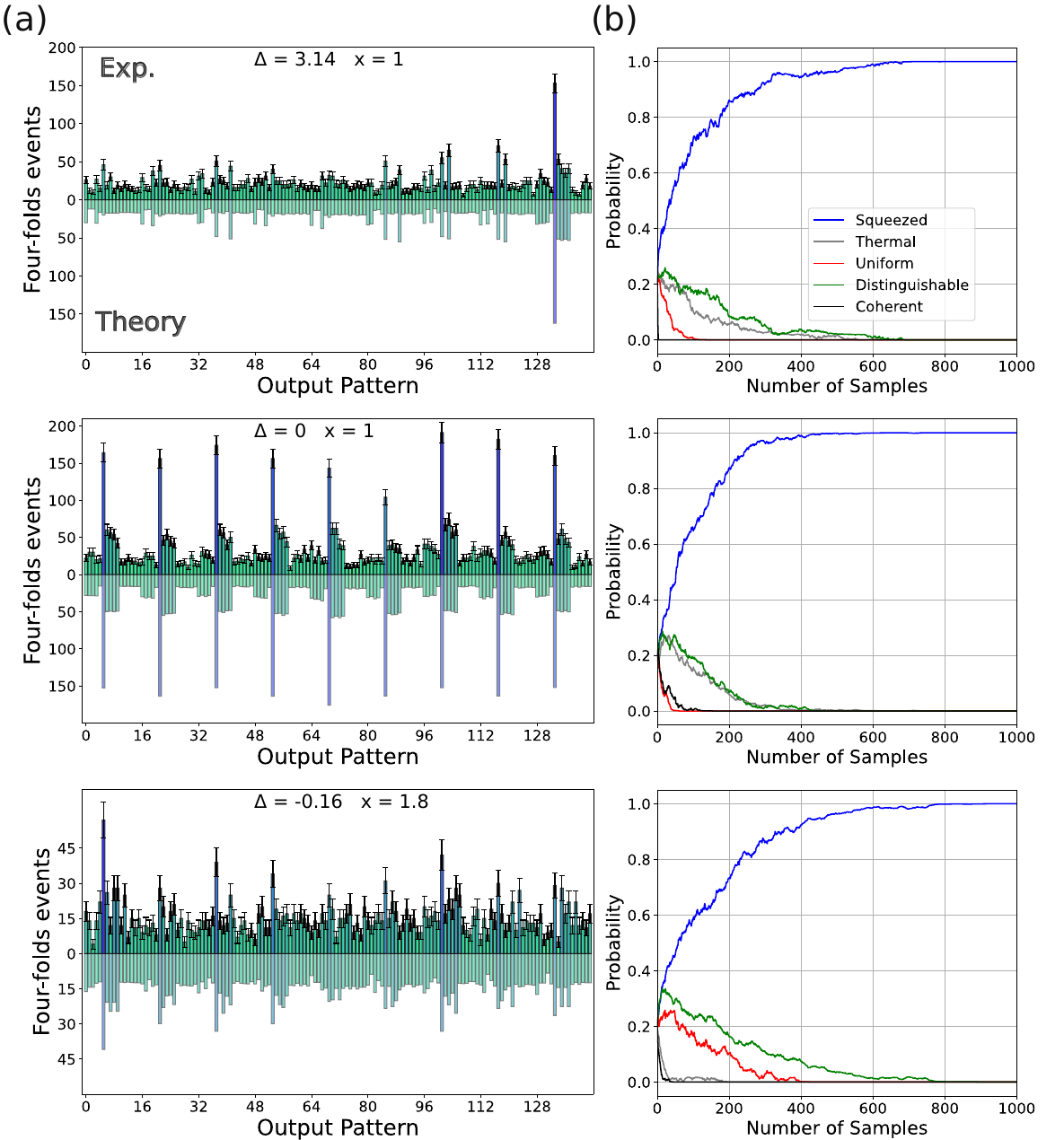}
    \caption{(a) Comparison between the experimental and the theoretical predicted number of four-photon events for $144$ different output patterns. The three stacked panels differ by the value of the phases $(x,\Delta)$. The mapping between the output pattern and the corresponding set of modes is indicated in the supplemental document. (b) Posterior probability that the collected dataset is generated by alternative models than squeezed light at the input of the interferometer as a function of the number of samples in the dataset.  
    }
    \label{Fig_1}
\end{figure*}
The phase $\Delta$ can be determined by measuring two-fold coincidences between the signal and idler photons at frequencies $[j,k]=[\nu_{s,1}+(j-1)\Delta\nu,\nu_{i,1}+(k-1)\Delta\nu]$ and detected in the late time-bin. From Eq.(\ref{Haf_to_perm_form}) and Eq.(\ref{eq:block_CLL}) the coincidence probability $p_{jk}$ is given by
\begin{equation}
    p_{jk} \propto 1+\cos\left(x+2\left ( \frac{j+k}{2}-1\right )\Delta\right ). \label{eq:cos_fit}
\end{equation}
Scanning $x$ yields an interference fringe with an offset that is a multiple of $\Delta$. There are $\binom{4}{2}=6$ possible combinations of signal and idler frequencies, which are $([0,0],[0,1],[0,2],[1,1],[1,2],[2,2])$, with respective offsets $(-2\Delta,-\Delta,0,\Delta,2\Delta)$. Equation (\ref{eq:cos_fit}) can be interpreted as a generalization of the more familiar Franson-type interference, in which photons are allowed to change their frequency. The interference arise from the pairs which are generated in the early pump pulse and travel the long path of the interferometer, and the pairs generated in the late pulse which travel the short path. Even if the PMs change the frequency of the signal or idler photon, these events are still indistinguishable and their amplitude probabilities interfere.
In Fig.\ref{Fig_Franson} we report an example of  measurement of the interference fringes for all the six combinations of signal-idler frequencies. All curves are fit using Eq.(\ref{eq:cos_fit}), leaving  $\Delta$ as a free parameter.  
As expected, there is no offset between the curves relative to the frequency combinations $[1,1]$ and $[0,2]$. On the other hand, the combinations $[0,0]$ and $[0,1]$ have offsets $-2\Delta$ and $-\Delta$, while the combinations $[1,2]$ and $[2,2]$ have offsets $\Delta$ and $2\Delta$.\\
After the characterization of the interferometer, we sampled the $144$ four-photon patterns at the output for four different values of $x$ and $\Delta$, and compared the number of collected events to those predicted by Eq.(\ref{Haf_to_perm_form}). The experiments are performed by setting the average number of photons per pulse to $\sim 0.1$, corresponding to a squeezing parameter of $|\xi| \sim 0.3$. The total integration time for each setting $(x,\Delta)$ is $13$ hours. The measured and predicted output patterns are shown in  Fig.\ref{Fig_1}(a) and are in very good agreement. The fidelity $\mathcal{F}$ between the theoretical  and experimental probability distributions $\bm{p}^{\textrm{th}}$ and $\bm{p}^{\textrm{exp}}$, defined as 
\begin{equation}
    \mathcal{F} = \sum_i \sqrt{p_i^{(\textrm{th})} p_i^{(\textrm{exp})}}, \label{eq:fidelity}
\end{equation} 
is $0.989(2)$ for $(x=1,\Delta=\pi)$, $0.992(1)$ for $(x=1,\Delta=0)$ and $0.987(2)$ for $(x=1.8,\Delta= -0.16)$.
Furthermore, we used a Bayesian approach to evaluate the posterior probability that the collected dataset is generated by alternative models than squeezed light at the input of the interferometer\cite{Sampling_Paesani}. We consider three possible alternative models of input state, which are thermal states, coherent states and distinguishable TMS. We also compare the output statistics to that predicted by uniform sampling. Details on the model validation procedure are reported in the Methods section. The posterior probabilities as a function of the number of samples in the dataset are shown in Fig.\ref{Fig_1}(b). For each pair of $(x,\Delta)$, the probability that the measured samples arise from any of the alternative models vanish after $\sim 10^3$ samples. This shows that the output samples are actually generated by squeezed light at the input of the interferometer.

\section{Evaluation of graph similarity}
\label{sec:graph_similarity}
It is known that a graph can be encoded in a GBS experiment through a correspondence
between the graph’s adjacency matrix and the combination of an interferometer with squeezed light \cite{BGS_orbite,GBS_application,Graph_opt_1}.
The graphs that can be encoded in this experiment are bipartite and represented by the adjacency matrix $\mathcal{A}$ in Eq.(\ref{eq:adj_matrix}), where the two vertex sets $V_s=(1,...,6)$ and $V_i=(1,...,6)$ are the signal and the idler time-frequency modes. The complex weight of an edge from a signal vertex $j$ to an idler vertex $k$ is given by the entry $C_{jk}$ of the matrix $C$ in Eq.(\ref{eq:Cmatrix}). 
\begin{figure*}[t!]
    \includegraphics[width = 1 \textwidth]{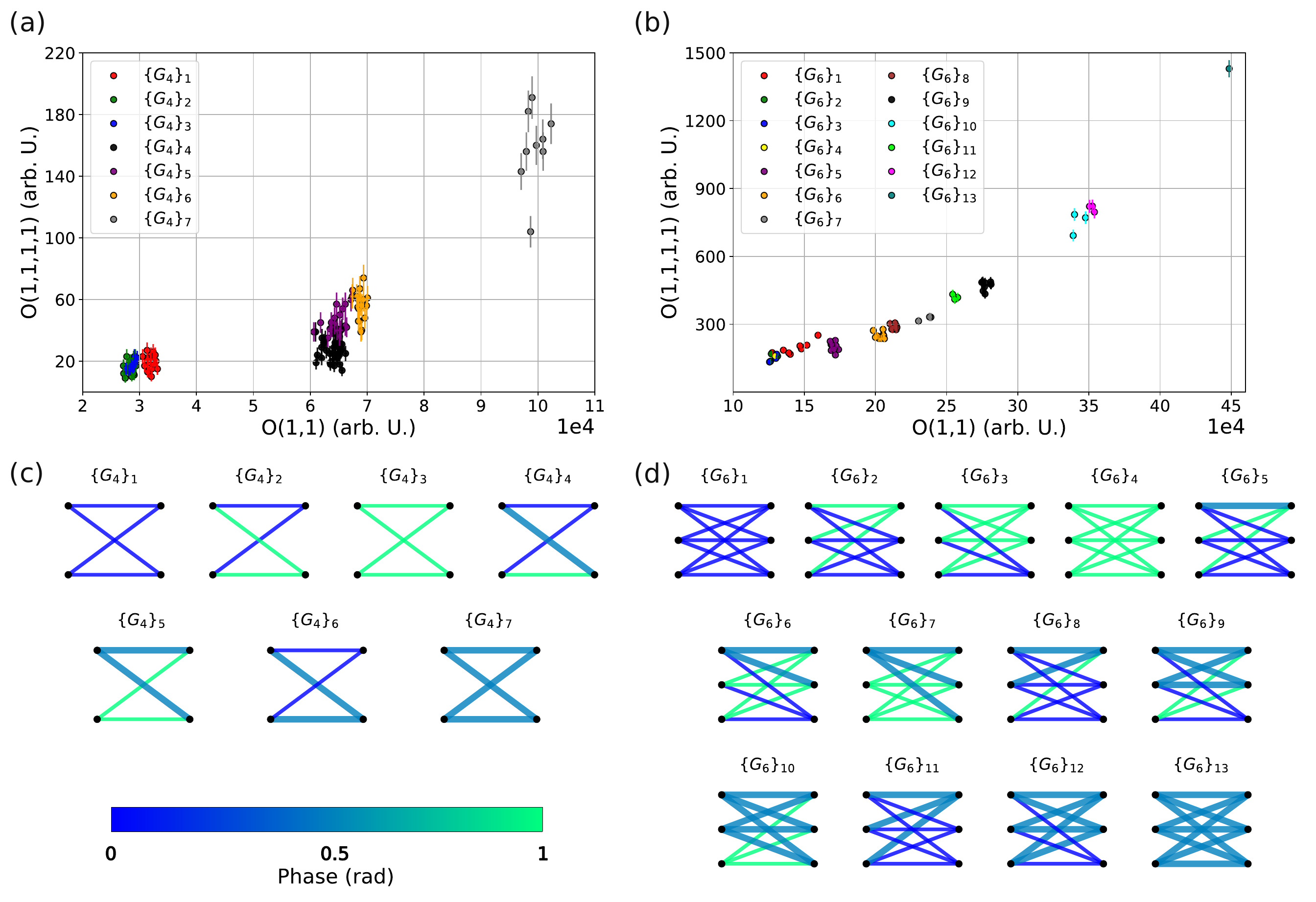}
    \caption{(a) Feature vectors corresponding to 7 different families of graphs with four vertices. The components of the vectors are the number of samples of the two and the four photons orbits $O_2$ and $O_4$. 
    (b) Same as in (a), but relative to 13 different families of graphs with six vertices. (c) Representation of the $7$ different families of graphs with $4$ vertices. The width of the edges represents the module of the complex weight, while the color encodes its phase. (d) Same as in (c), but relative to the $13$ families of $6$ vertex graphs.
    }
    \label{Fig_graphs}
\end{figure*}
Given two GBS experiments encoding isomorphic graphs, any output pattern $\bm{n}$ has the same probability to occur as a pattern $\bm{n}'$ that is related to $\bm{n}$ by a permutation of the modes \cite{GBS_application}. Hence, one can identify if two graphs are isomorphic by comparing the output probabilities of the two experiments. The problem of determining if two graphs are isomorphic problem falls in the NP complexity class, which motivates the use of GBS for providing a computational speed-up over classical algorithms \cite{GBS_application}. The problem of estimating a number of output probabilities that grows combinatorially with the number of vertices has been tackled in refs. \cite{GBS_application,arrazola2021quantum}, where the concepts of orbits and feature vectors on graphs are implemented. An orbit is defined as a set of output patterns that are equivalent under permutation. For example, the orbits $[1,1,1,1]$ and $[1,1]$ contain all the detection events where $4$ photons and $2$ photons are respectively detected in separate modes, without distinguishing the mode number, and with zero photons in all the other modes. Two GBS experiments encoding isomorphic graphs have identical probabilities of generating samples from the same orbit, which suggest that orbits can be combined into feature vectors whose distance depends on the degree of similarity between two graphs.\\ 
In our implementation, we set $x=1$ and $\Delta=0$ and collected the output samples with $2$ and $4$ photons. 
From the complete bipartite graph with $12$ vertices, we extracted $7$ different families of graphs with $4$ vertices $\{G_4\}_{i=1,...,7}$, and $13$ different families with $6$ vertices $\{G_6\}_{i=1,...,13}$, which are shown in Fig.\ref{Fig_graphs}(c,d). Each family is formed by isomorphic graphs, but the graphs belonging to different families are not isomorphic. The adjacency matrix of a representative graph within each family and the bijective functions between the vertex sets of isomorphic graphs are reported in the supplemental document.  We clustered the recorded events into two and four-photon orbits $O_2$ and $O_4$. More specifically, these orbits are formed by two and four-photon events while tracing out the unobserved modes, which are therefore not post-selected to contain zero photons. However, we will still refer to $O_2$ and $O_4$ as orbits that can be used to construct feature vectors for the purpose of graph clustering and classification. We assign to each graph its feature vector $\bm{f}=(N(O_2),N(O_4))$, where $N(O_i)$ is the number of samples in the orbit $O_i$, and plot the feature vectors in Fig.\ref{Fig_graphs}(a). We see that the isomorphic graphs within each family have very similar feature vectors, thereby forming clusters. We can clearly recognize $6$ separate clusters of graphs in Fig.\ref{Fig_graphs}(a), and $11$ in Fig.\ref{Fig_graphs}(b), corresponding to different families of non-isomorphic graphs.
Therefore, the approximate orbits can be used to identify isomorphic graphs in our small scale problem. 
The overlapped groups (e.g., $\{G_4\}_2$ and $\{G_4\}_3$ for $4$ vertex graphs and $\{G_6\}_2$ and $\{G_6\}_3$ for six vertex graphs)  could be separated in principle by adding more and/or different orbits to the feature vector.
 

\section{Universal operations on time-frequency modes}
\begin{figure*}[t!]
    \includegraphics[width = 1 \textwidth]{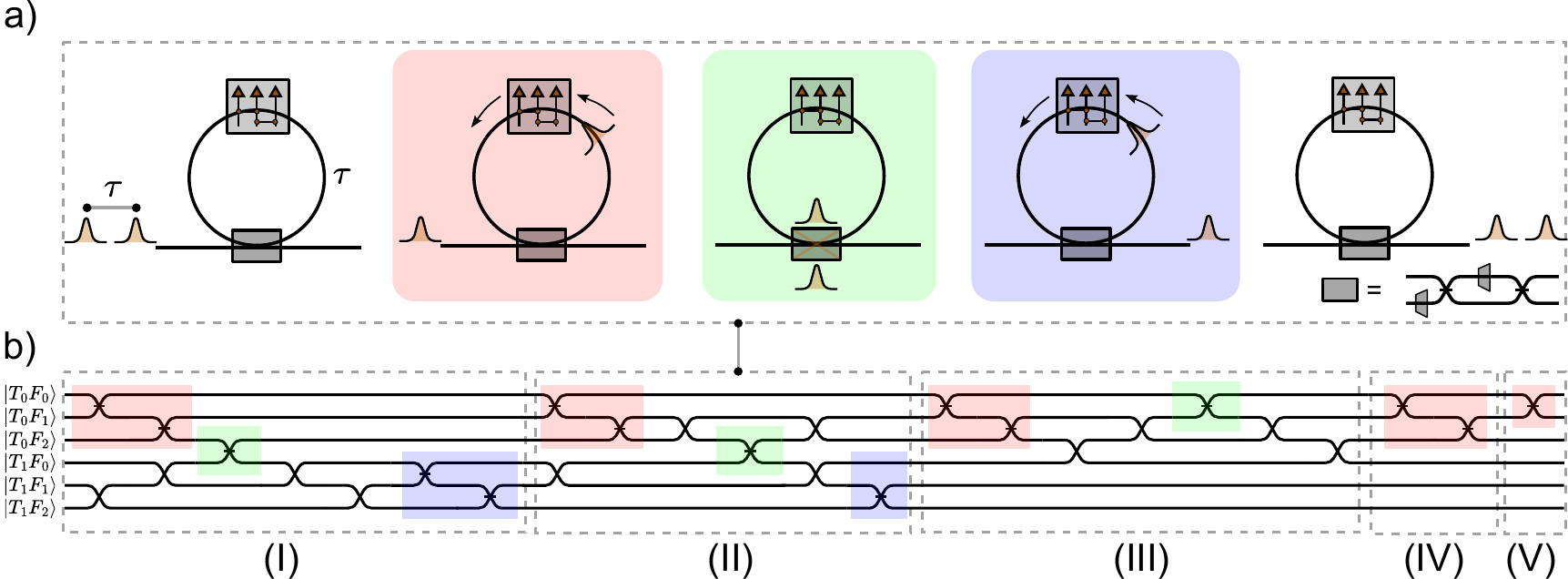}
    \caption{(a) The sequence of operations in the time and frequency domain executed in each of the five stages of the universal interferometer. From left to right: the first pulse, separated by a time $\tau$ from the second, is coupled to the loop and passes through a QFP, which perform a unitary operation over the frequency-bins (red). The two pulses met at the beamsplitter (green), mixing the two time-bins. The reflected pulse propagates in the loop, passes in the QFP where it undergoes a unitary transformation over the frequncy-bins (blue). At the end of this operation, the pulse exits the loop. The bottom-right inset shows a possible implementation of the frequency-dependent beamsplitter. The element within the MZI is a waveshaper. (b) Equivalent decomposition in terms of $2\times2$ beamsplitters and mode-swaps of the universal interferometer over six time-frequency modes $\ket{T_iF_j}$ ($i=\{0,1\}$, $j=\{0,1,2\}$). The order of the operations in each stage follows the steps indicated from left to right in panel (a). Each stage is bounded by a dashed box. The physical implementation of the beamsplitter operation is indicated by the panel of the same color in (a).    
    }
    \label{Fig_4}
\end{figure*}
The transformation applied to the time-frequency modes by the circuit in Fig.\ref{Fig_0}(a) is achieved by using a simple experimental setup, but it is not  universal. In this section we show that a universal interferometer over $N$ modes can be realized by a series of $N-1$ elementary building blocks made by a combination of waveshapers, beamsplitters and electro-optic modulators. It has been shown in \cite{motes2014scalable} that a scalable and universal architecture in the time-bin domain can be built by using a series of $N-1$ fiber-loops and beasmplitters whose splitting-ratio can be reconfigured every round-trip time. The proof is based on the observation that the sequence of operations can be mapped into a universal Reck-style decomposition of a general unitary matrix \cite{reck1994experimental}.  In parallel, e recipe for realizing a universal transformation over frequency-bin modes has been given in \cite{lukens2016frequency} in terms of an alternating sequence of electro-optic modulators and waveshapers. This fundamental unit is generically termed the quantum frequency processor (QFP) \cite{lu2023characterization}. 
In this case, the proof builds upon the factoring of a $N\times N$ complex matrix as the product of no more than $2N-1$ circulant and diagonal matrices, or equivalently, of $2N-1$ diagonal matrices spaced by discrete Fourier transform matrices. These transformations are exactly those implemented by waveshapers and electro-optic modulators. It is worth mentioning that while $2N-1$ is an upper bound to the number of required optical elements, the problem of finding the minimum number of resources is still open and currently relies on numeric optimization \cite{lu2022high,henry2023parallelizable}.\\
It is therefore  not surprising that a universal architecture over $N$ time-frequency modes can be built by combining the two strategies. In particular, we propose a multi-stage architecture, where the building block shown in Fig.\ref{Fig_4}(a) is repeated $N-1$ times. Within each block, light pulses can be coupled into a fiber-loop through a frequency-dependent beamsplitter that is reconfigured at each round-trip. For example, this element could be realized by a balanced MZI in which a waveshaper is placed in one of the arms, thus imparting a frequency-dependent phase. The loop introduces a delay $\tau$ which matches the time separation between two adjacent temporal modes. A QFP is placed inside the loop to perform an arbitrary unitary operation over the frequency-bin modes, and is reprogrammed every round-trip time. In the Methods section, we show that each stage implements two fundamental operations: nearest-neighbor coupling of the temporal-modes and arbitrary mixing of the frequency modes, including mode swaps. These operations are sufficient to realize a universal linear transformation over $N$ time-frequency modes, because they can be repeatedly applied to implement a Gaussian elimination of any unitary matrix $U$ into a row-echelon form \cite{reck1994experimental}. As an example, Fig.\ref{Fig_4}(b) shows the sequence of beamsplitter operations performed by each of the $N-1$ stages which implement an arbitrary transformation over six time-frequency modes $\ket{T_iF_j}$ ($i=\{0,1\}$, $j=\{0,1,2\})$ (see Methods section for a detailed description of each stage).
Similarly to the fiber-loop based architecture for time-bins \cite{motes2014scalable}, there is no need to build $N-1$ physically distinct stages, because each of them requires exactly the same physical resources, only differing by the programmable beamsplitter splitting-ratio and the settings of the QFP. A single loop can then be embedded into a larger fiber loop of delay $>N\tau$ by using two switches \cite{motes2014scalable}.
The interferometer in Fig.\ref{Fig_0}(a) can be seen as a simple instance of a single stage of the universal scheme of Fig.\ref{Fig_4}(a). In particular, the beasmplitter used to couple the time-modes has a fixed splitting ratio of $50/50$ and is not frequency-dependent, while the QFP is replaced by a single electro-optic phase modulator. 
By construction, the architecture in Fig.\ref{Fig_4}(a) can implement any linear operation over time-frequency modes, but no efforts have been made to reduce the number of physical resources. Therefore, more efficient strategies may exist, whose investigation lies outside the scope of this manuscript. However, we recognize that the implementation of this architecture is currently very challenging because of the losses associated to the QFP \cite{henry2023parallelizable} and the need to reconfigure the QFP and the frequency-dependent beamsplitter at high-speed. Nevertheless, on chip-waveshapers with GHz resolution have been recently reported \cite{cohen2024silicon}, and there is hope that hybrid-platforms embedding electro-optic phase modulators \cite{wang2023integrated}, or new paradigms for frequency manipulation \cite{buddhiraju2021arbitrary,cui2020high} could decrease the loss of aggregate components.

\section{Conclusions}
We performed multi-photon quantum interference in the frequency and time-bin domain between multiple pairs generated by a silicon nitride microresonator. The device generates bichromatic two-mode squeezing in two temporal modes with simultaneous high spectral purity and indistinguishability. We sampled four-photon events at the output of a fiber-based interferometer acting both on time and frequency modes. Frequency mixing is achieved by electro-optic phase modulation, while the coupling between temporal modes is achieved by an unbalanced interferometer that is stabilized by exploiting the time-energy entanglement of the photon pairs. The output patterns have a high fidelity with those predicted by simulations, and can not be explained by feeding classical light at the input of the interferometer. We used the connection between GBS and graphs to cluster families of isomorphic graphs and identify the similarity between groups of not-isomorphic graphs. Lastly, we proposed an architecture to perform arbitrary operations on time-frequency modes.
This work demonstrates that time and frequency-bin encoding can be combined together to increase the dimensionality of sampling experiments, which can be performed in a single spatial mode and using standard fiber-optic telecommunication components. We anticipate that the integration of low-loss quantum frequency processors on-chip will enable universal operations and further scale the size of experiments.    

\section*{Funding and acknowledgements}
D.B. acknowledges the support of Italian MUR and the European Union - Next Generation EU through the PRIN project number F53D23000550006 - SIGNED. E.B., M.B., M.G. and M.L. acknowledge the PNRR MUR project PE0000023-NQSTI. All the authors acknowledge the support of Xanadu Quantum Technology for providing the samples.
\section*{Methods}
\subsection{Description of the architecture for universal linear-transformation over time-frequency modes}
Here we prove that an arbitrary linear transformation over $N$ time-frequency modes can be realized by cascading $N-1$ fundamental building blocks, each implementing the sequence of operations shown in Fig.\ref{Fig_4}(a). As an example, we focus on an arbitrary transformation over six time-frequency modes $\ket{T_iF_j}$ ($i=\{0,1\}$, $j=\{0,1,2\})$. With reference to Fig.\ref{Fig_4}(a), the following steps are executed in sequence:
\begin{enumerate}
    \item The first pulse is coupled into the loop and passes through the QFP, which implements a unitary operation over the three frequency-bins $\ket{T_0F_j}\, ,j=\{0,1,2\}$ (highlighted in red). In the representation using spatial modes, shown in Fig.\ref{Fig_4}(b), this corresponds to a sequence of beasmplitter operations between modes $(0,1)$ and $(1,2)$ (highligted in red). The splitting ratio depend on the stage $n$. 
    \item The two pulses met at the beasmplitter (highlighted in green). The reflected pulse remains in the loop, while the transmitted pulse exits the loop and forms the first time-bin for the next stage. The splitting-ratio of the beasmplitter depend on frequency, and only two frequency modes $(\ket{T_0F_j},\ket{T_1F_j})$ are coupled. The modes which are coupled and the splitting ratio depend on the stage $n$.
    \item The reflected pulse passes through the QFP, which implements a unitary operation over the three frequency-bins $\ket{T_1F_j}\, ,j=\{0,1,2\}$ (highlighted in blue). In the representation using spatial modes, shown in Fig.\ref{Fig_4}(b), this corresponds to a sequence of beasmplitter operations between frequency-bin modes $(0,1)$ and $(1,2)$ (highligted in blue). The splitting ratios depend on the stage $n$. 
    \item The reflected pulse exits the loop, where it forms the second time-bin for the next stage.   
\end{enumerate}
After these steps, the whole procedure is repeated again in the next stage $n+1$. At the output of the last stage, the whole transformation will the target unitary matrix $U$. Similarly to the Reck-decomposition, the sequence of beasmplitter operations at stage $n$ is chosen to nullify the elements $U_{nj}$ with $j=(1,...,n-1)$. The gaussian-elimination pattern of the decompositon shown in Fig.\ref{Fig_4}(b) is
\begin{equation}
\left (  \begin{matrix}
\cdot & \cdot & \cdot & \cdot & \cdot & \cdot \\
15& \cdot & \cdot & \cdot & \cdot & \cdot \\
13 & 14 & \cdot& \cdot & \cdot & \cdot \\
12 & 10 & 11& \cdot & \cdot & \cdot \\
6 & 8 & 7& 9 & \cdot & \cdot \\
1 & 2 & 3& 4 & 5 & \cdot \\
\end{matrix} \right ),
\label{eq:U_decomposition}
\end{equation}
where the numbers indicate the order at which the corresponding matrix element is nullified. For example, the entry $U_{5,2}$ is nullified after $8$ beamsplitter operations. As it happens in Reck's decomposition \cite{reck1994experimental}, the pattern is chosen to prevent subsequent beamsplitter operations from affecting elements of $U$ that have been nullified by previous operations.    
We see from the pattern in Eq.(\ref{eq:U_decomposition}) that the modes $r=(1,..,N)$ and $s=(1,...,N)$ which are mixed by the frequency-dependent beasmplitter at stage $n$ are $r=3-(n-1)\,,s = 6-(n-1)$. This result can be generalized by inspection to $Q$ time-bins and $M$ frequency-bins, corresponding to $N=MQ$ modes. In this case, at stage $n$ the mixed modes will be $r=qM-n+1$ and $s=(q+1)M-n+1$, with $q=(1,...,Q-1)$. \\
\subsection{Impact of thermal noise and multi-photon contamination on the visibility of the heralded HOM interference}
\begin{figure}[b!]
    \includegraphics[width = 0.5 \textwidth]{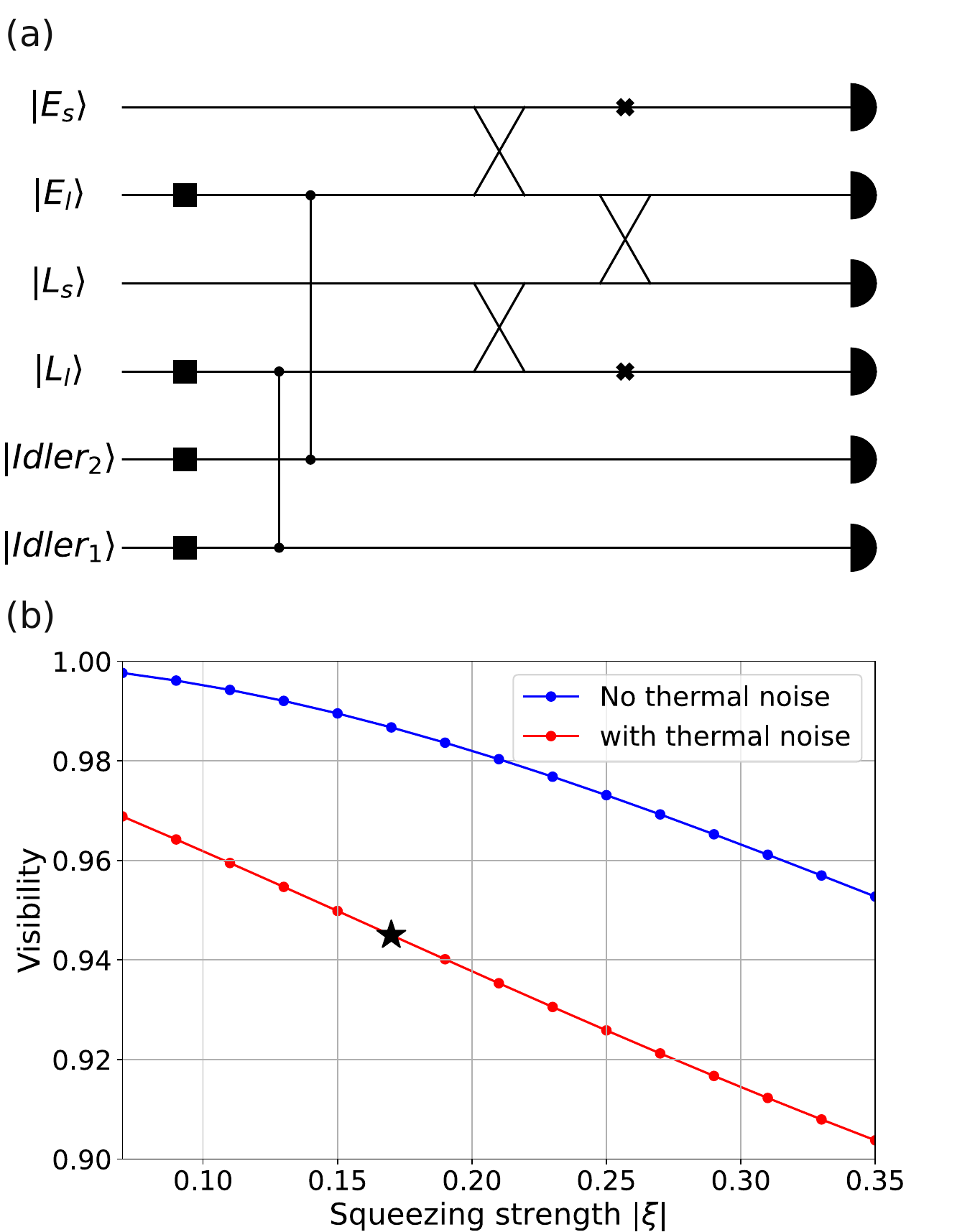}
    \caption{(a) Circuit for the simulation of the heralded Hong-Ou-Mandel experiment using \texttt{Strawberry Fields}. The black rectangle represents a thermal loss channel, the vertical line connecting two modes represents a TMS operation, the "x" represents a pure loss channel, the crossing between two modes represents a $50/50$ beamsplitter, and the black semi-circles represent threshold detectors. (b) Visibility of the HOM dip as a function of the squeezing strength $|\xi|$. The red line includes thermal noise photons on the squeezed modes. The black star indicates the squeezing strength used in the heralded HOM experiment reported in Fig.\ref{Fig_0}(d).}
    \label{HHOM_scheme}
\end{figure}
Here we study through numerical simulations the impact of multi-photon events and thermal noise to the visibility of the heralded HOM interference.
We assume that photons are heralded in a single spectral mode, and that photons heralded from different sources are indistinguishable. This allows us to isolate the effect of multi-photon contamination and of thermal noise from other factors that limit the visibility of the HOM dip \cite{Drago_HOM_24}.
The numerical simulations are performed using the open source software \texttt{Strawberry Fields} \cite{Killoran2019strawberryfields}. The gaussian backend
(GB) of the software allows handle any gaussian operation, which
include thermal-loss channels. Furthermore, one can use
the plug-in library \texttt{The Walrus} to simulate
threshold detection. We used the circuit shown in Fig.\ref{HHOM_scheme}(a) to calculate the visibility. Two TMS, which in our experiment are generated by the two consecutive pump pulses pulses, are initialized in the pairs of mode $\ket{E_l}-\ket{\textrm{Idler}_2}$ and $\ket{L_l}-\ket{\textrm{Idler}_1}$. All these channels are subjected to a thermal-loss operation that simulates losses from generation to detection and the injection of thermal noise, which in our case is induced by the spontaneous Raman scattering generated on-chip and in the fibers \cite{brusaschi2024photon}.
The idler modes are detected and act as heralding events for the signal photons. To simulate the on/off detection, we used the \texttt{threshold\textunderscore detection\textunderscore prob} module of \texttt{The Walrus} library.
The modes $\ket{E_l}$ and $\ket{L_l}$ are then fed into separate $50/50$ beamsplitters together with vacuum states, which simulate first physical beamsplitter of the inferometer in Fig.\ref{Fig_0}(a). After that, photons have a 50/50 probability to travel the long or the short path of the interferometer. This defines four modes, that are labeled $\ket{E_s}$,$\ket{E_l}$,$\ket{L_s}$,$\ket{L_l}$, where $E(L)$ indicates photons which are generated in the early/late) pulse, while the subscript $s(l)$ refer to photons that travel the short(long) arm of the interferometer.
The modes $\ket{E_l}$ and $\ket{L_s}$ are combined into a second $50/50$ beamsplitter, which emulates the action of the delay $\tau$ and of the second beasmplitter in the interferometer shown in Fig.\ref{Fig_0}(a). Coincidence detection is performed at the output ports using threshold detectors to count photons that arrive simultaneously. In the ideal case of truly single photons at the input and without thermal noise, the number of observed coincidences $N_0$ should be zero.
To calculate the visibility of the HOM dip we need to simulate the number of coincidences $N_{td}$ between temporally distinguishable photons at the output ports of the beasmplitter.
We simulate the non-interfering photons using modes $E_s$ and $L_l$, representing photons generated in the early pulse that travel the short arm of the interferometer, and photons generated in the late pulse that travel the long arm. These photons arrive at the second beasmplitter at different times and do not interfere. This beasmplitter is then equivalent to a loss channel (see Fig.\ref{HHOM_scheme}), where half of the photons are lost. The number of coincidence detection at the output modes $E_s$ and $L_l$ is then $N_{td}$, which allows one to calculate the visibility of the HOM dip as 
\begin{equation}
    V = 1 -\frac{N_0}{N_{td}} \label{eq:hom_visibility}.
\end{equation}
We used $15$ dB of loss for signal photons and $10$ dB of loss for idler photons, which correspond to the experimental values. In the simulation, the squeezing parameter $\xi$ is varied from $0.07$ to $0.35$.
The average number of thermal photons $\langle n_{\textup{th}}\rangle$ injected into the circuit of Fig.\ref{HHOM_scheme} is assumed to linearly grow with the squeezing parameter. This assumption is justified by the fact that for moderately low squeezing levels $\xi\propto P$ \cite{brusaschi2024photon}, where $P$ is the pump power, and the level of thermal noise also linearly grows with $P$ \cite{brusaschi2024photon,borghi2024uncorrelated}. Using the strategy outlined in ref. \cite{brusaschi2024photon}, we experimentally determined that $\langle n_{\textup{th}}\rangle = k|\xi|$, with $k=0.1$.
The result of the simulation is shown in Fig.\ref{HHOM_scheme}(b). The visibility decreases as $\xi$ and $\langle n_{\textup{th}}\rangle$ increase, as expected. In the experimental conditions of the HOM experiment, the squeezing parameter is set to
$|\xi|=0.17$ and an the average number of thermal photons is estimated to be $\langle n_{\textup{th}}\rangle=0.017$. The visibility that can be observed in the experiment is then upper bounded to $0.945$ by multi-photon contamination and thermal noise. 
\subsection{On the use of the permanent for predicting the output pattern probabilities}
\label{subsec:on_the_use_of_the_permanent}
\begin{figure}[t!]
    \includegraphics[width = 0.45 \textwidth]{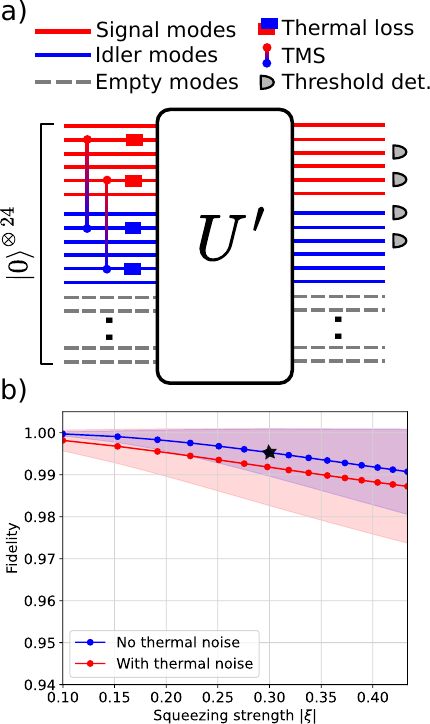}
    \caption{Sketch of the circuit which is simulated using the GB kernel. (b) Fidelity between the output probability distribution calculated using Eq.(\ref{Haf_to_perm_form}) and the GB kernel as a function of the squeezing strength $|\xi|$. The shaded regions represent the standard deviation over $400$ different realizations of $x$ and $\Delta$, sampled uniformly between $[-\pi,\pi]$. The black star indicates the squeezing strength in the experiment.}
    \label{Fig:SF_circuit}
\end{figure}
The probability of observing a particular output pattern $\ket{S}$ can be calculated by using Eq.(\ref{Haf_to_perm_form}) from the matrix $C$ given in Eq.(\ref{eq:Cmatrix_block}). We used this in Section \ref{sec:graph_similarity} to link the output distributions to weighted graphs and the graph isomorphism problem. However, Eq.(\ref{Haf_to_perm_form}) strictly holds only when the modes $S'\neq S$ are post-selected to contain zero photons and lossless photon number resolving detection is performed on the $S$ modes \cite{kruse2019detailed,arrazola2021quantum}. These requirements are not satisfied in our experiment. Indeed,  we use threshold detectors to monitor the lossy $S$ modes, and we leave the $S'$ modes undetected. In addition, we have thermal noise generated by spontaneous Raman scattering that alters the output statistics. In order to validate the distance between the outcome probabilities predicted by Eq.(\ref{Haf_to_perm_form}) and those in the real experimental conditions, we used the open source software \texttt{Strawberry Fields}. 
Our experiment is modeled using the equivalent circuit shown in Fig.\ref{Fig:SF_circuit}(a). The six signal and idler modes are initialized in vacuum, and two-mode squeezing is performed over the set of modes $\ket{E_1}_s-\ket{E_1}_i$ and $\ket{L_1}_s-\ket{L_1}_i$. Losses and thermal noise are implemented via thermal-loss channels on these modes. Then, the modes enter the interferometer $U'$, which is described by a unitary matrix dilation of the physical interferometer $T_{si}=T_s\bigoplus T_i$ acting on the signal-idler modes, where $T_{s(i)}$ is given by Eq.(\ref{Matrice_idler}) (the irrelevant entries denoted with a star are set to zero). We choose the following dilation of $U'$:
\begin{equation}
    U' = 
    \begin{bmatrix}
    T_{si}/\sigma_{\textup{max},\epsilon} & R\sqrt{\mathbb{I}-D^2}V \\
    R\sqrt{\mathbb{I}-D^2}V & -T_{si}/\sigma_{\textup{max},\epsilon} 
    \end{bmatrix}, \label{eq:matrix_dilation}
\end{equation}
where $T_{si}=RDV$ is the singular value decompositon of $U$, $D$ is a diagonal  matrix filled with the eigenvalues $\bm{\sigma}$ of $T_{si}$ and $\sigma_{\textup{max},\epsilon} = \textrm{max}(\bm{\sigma})+\epsilon$ ($\epsilon = 10^{-3}$ is a regularization parameter which makes all the eigenvalues of $T_{si}/\sigma_{\textrm{max},\epsilon}$ to be $\leq 1$). The procedure of unitary dilation is required by the GB kernel of SF because $T_{si}$ is not unitary. The program then runs over a doubled number of modes. The first half of the modes correspond to the physical modes which are coupled by the time-frequency interferometer $T_{si}$, while the second half are just auxiliary modes which are all initialized in vacuum. To simulate the on/off detection at the output of the interferometer, we used the \texttt{threshold\textunderscore detection\textunderscore prob} module of \texttt{The Walrus} library, which use the formulas in \cite{bulmer2022threshold} to calculate the click probabilities of multi-photon patterns from a gaussian state described by a covariance matrix $\Omega$ and zero displacement by using threshold detectors. To account for the undetected modes $S'$, the covariance matrix at the output of the interferometer is reduced to include only the modes $S$, which is equivalent to a partial trace operation over the modes $S'$ of the system density matrix. Figure \ref{Fig:SF_circuit}(b) compares the fidelity $\mathcal{F}$ (see Eq.(\ref{eq:fidelity})) between the output probability distribution predicted using Eq.(\ref{Haf_to_perm_form}) and the GB kernel for an increasing squeezing strength $|\xi|$. For each value of $\xi$, we evaluated $\mathcal{F}$ over $400$ different combinations of $x$ and $\Delta$, sampled uniformly between $[-\pi,\pi]$. As expected, the fidelity decreases as $|\xi|$ increases  because the simultaneous emission of more than four photons photons becomes more likely. However, $\mathcal{F}>0.99$ for $|\xi|\le 0.45$, and $\mathcal{F}=0.995(5)$ at $|\xi|=0.3$, which is the squeezing strength used in the experiment. Therefore, despite the use of threshold detectors and the trace operation over the undetected modes, Eq.(\ref{Haf_to_perm_form}) approximates well the true output probability. We also simulated the impact of thermal noise on the output statistics. As shown in Fig.\ref{Fig:SF_circuit}, the addition of thermal noise lowers the value of $\mathcal{F}$ for any squeezing level, but the reduction is not significative ($\mathcal{F}$ reduces from $0.995$ to $0.991$ at the squeezing strength adopted in the experiment).
\subsection{Bayesan model comparison}
Bayesan approaches have been widely implemented to compare an
ideal experimental implementation of GBS against a general test models for which output
probabilities are computable \cite{Sampling_Paesani}. The idea is to consider a set $\mathcal{M}_{1,..,M}$ of plausible models describing the experiment and calculate the probability $p(D|\mathcal{M})$ to observe the experimental dataset $D$. Using Bayes's rule, one can then calculate the probability $p(\mathcal{M}_i|D)$ that the model $\mathcal{M}_i$ describes the experiment given the dataset $D$ as
\begin{equation}
    p(\mathcal{M}_i|D) = \frac{p(D|\mathcal{M}_i)p(\mathcal{M}_i)}{\sum_i^M p(D|\mathcal{M}_i)},\label{eq:bayes_rule}
\end{equation}
where $p(\mathcal{M}_i)$ is the prior probability attribuited to the model $\mathcal{M}_i$, which in our case is assumed to be uniform ($\mathcal{M}_i=1/M$). We compare the following models: indistinguishable two-mode squeezed states at the input, distinguishable squeezers at the input, thermal state and coherent state at the input and uniform sampling. For indistinguishable squeezers, thermal and coherent states at the input, the probability $p(D|\mathcal{M})$ is calculated using the GB kernel. We simply replaced the squeezed modes at the input shown in Fig.\ref{Fig:SF_circuit}(a) with thermal and coherent states (both gaussian states) having the same average photon number. For the uniform sampler, $p(D|\mathcal{M})$ is simply given by $\left (\frac{1}{144} \right)^q$, where $q$ is the cardinality of the set $D$. In the case of distinguishable squeezers at the input, the probability of detecting four photons at the output of the interferometer in a specific pattern $S=(m_s,n_s,m_i,n_i)$ (with $m_{s(i)},n_{s(i)} = 1,...,6$ respectively the signal and the idler modes) is calculated as $p_{40}(s)+p_{04}(s)+p_{22}(s)$, where $p_{ij}(s)$ is the probability to detect $i+j$ photons at the output of the interferometer in pattern $S$ given that $i$ photons are generated by the first source (early pulse) and $j$ photons by the second source (late pulse). By using the same numbering of modes introduced in Section \ref{sec:multi-photon_interference}, the three probabilities are given by
\begin{equation}
    p_{40}(s) = |\textrm{perm}([T_s]_{1,1,s_1,s_2})\textrm{perm}([T_i]_{1,1,s_3,s_4})|^2,
\end{equation}
\begin{equation}
    p_{04}(s) = |\textrm{perm}([T_s]_{5,5,s_1,s_2})\textrm{perm}([T_i]_{5,5,s_3,s_4})|^2,
\end{equation}
\begin{equation}
    p_{22}(s) = |\textrm{perm}([T_s]_{1,5,s_1,s_2})\textrm{perm}([T_i]_{1,5,s_3,s_4})|^2.
\end{equation}
The matrices $[T_{s(i)}]_{i,j,s_k,s_r}$ are obtained from $T_{s(i)}$ by selecting the rows $(i,j)$ and the columns $(s_k,s_r)$.

%

\end{document}